\documentclass[11pt]{article}
\usepackage{amsmath,amssymb,epsfig,bm}

\date{\empty}

\begin{document}

\title{\bf Gravito-magnetic amplification in cosmology}

\author{Christos G. Tsagas\\ {\small Section of Astrophysics, Astronomy and Mechanics, Department of Physics}\\ {\small Aristotle University of Thessaloniki, Thessaloniki 54124, Greece}}

\maketitle

\begin{abstract}
Magnetic fields interact with gravitational waves in various ways. We consider the coupling between the Weyl and the Maxwell fields in cosmology and study the effects of the former on the latter. The approach is fully analytical and the results are gauge-invariant. We show that the nature and the outcome of the gravito-magnetic interaction depends on the electric properties of the cosmic medium. When the conductivity is high, gravitational waves reduce the standard (adiabatic) decay rate of the $B$-field, leading to its superadiabatic amplification. In poorly conductive environments, on the other hand, Weyl-curvature distortions can result into the resonant amplification of large-scale cosmological magnetic fields. Driven by the gravitational waves, these $B$-fields oscillate with an amplitude that is found to diverge when the wavelengths of the two sources coincide. We present technical and physical aspects of the gravito-magnetic interaction and discuss its potential implications.\\\\PACS numbers: 98.80.Jk, 04.30.-w, 98.62.En
\end{abstract}

\section{Introduction}\label{sI}
The origin of the large-scale magnetic fields seen in the universe today remains an enigma, despite the efforts of the scientific community to resolve it~\cite{KWM,E}. The structure of the galactic $B$-fields, primarily of those found in spiral galaxies, seems to support the dynamo idea~\cite{P}. However, the galactic dynamo requires an initial seed-field to operate and the origin of these magnetic seeds is still elusive. Provided the amplification is efficient, the strength of the seed can be as low as $\sim10^{-23}$~G by the time the host galaxy is formed~\cite{K}. This requirement could be relaxed down to $\sim10^{-34}$~G in dark-energy dominated universes~\cite{DLT}. In the absence of dynamo, seeds close to $10^{-12}$~G (or even up to $10^{-8}$~G) are necessary to meet the observations. An additional issue is the scale of the initial magnetic field, since the galactic dynamo requires seeds with a minimum (comoving) coherence length of $\sim10$~Kpc to operate successfully~\cite{P}.\footnote{All scales and strengths have been redshifted to today, unless stated otherwise.}

The literature contains various scenarios of cosmic magnetogenesis, which are traditionally clasified into those operating after recombination and those advocating a cosmological origin for the large-scale magnetic fields of the universe. Primordial magnetism is an attractive idea because it makes these large-scale fields  (especially those found in distant proto-galactic structures) easier to explain. The case was strengthened further when recent observations detected substantially strong magnetic fields at very high redshifts~\cite{KBMLSH}. Nevertheless, early-universe magnetogenesis is still not problem-free. Producing magnetic seeds that satisfy the above mentioned dynamo requirements has, as yet, proved difficult. There are problems both with the scale and the strength of the initial field. Causality means that all magnetic seeds generated between inflation and (roughly) recombination have coherence lengths well below the minimum dynamo requirements. For instance, $B$-fields produced during the electroweak phase transition span scales of the order of the astronomical unit. A mechanism known as `inverse cascade' can increase the correlation length by transferring magnetic energy to larger scales, but typically requires large-amounts of helicity~\cite{BEO}. Inflation has long been considered as an answer to the scale problem, since it naturally produces long-wavelength correlations. However, $B$-fields that have survived a phase of de Sitter expansion are typically too weak to sustain the galactic dynamo, unless classical electromagnetism is modified~\cite{TW,MS-J}, or Friedmann-Robertson-Walker (FRW) hosts with non-Euclidean spatial geometry are employed~\cite{TK}.\footnote{Large-scale magnetic fields in Friedmann universes have long been believed to decay `adiabatically'. The conformal flatness of the FRW models and the conformal invariance of Maxwell's equations are thought to guarantee a Minkowski-like ($B\propto a^{-2}$, with $a$ being the cosmological scale factor) decay-rate for the $B$-field. However, although all Friedmann universes are conformally flat, only the spatially flat is globally conformal to Minkowski space. For the rest, the conformal mappings are local. As a result, the adiabatic magnetic decay is only guaranteed in FRW models with Euclidean spatial sections. Magnetic fields embedded in Friedmannian universes with open spatial sections are `superadiabatically' amplified on scales close to the curvature radius~\cite{TK}. The hyperbolic geometry slows the decay rate to $a^{-1}$, leading to astrophysically interesting residual $B$-fields without breaking away from conventional electromagnetism or abandoning the symmetries of the FRW spacetimes.}

Gravitational waves could also provide a geometrical mechanism of magnetic amplification on cosmological scales. Gravitons are readily produced in almost all inflationary scenarios and interact with magnetic fields through the shear anisotropy that they induce. The  theory behind the coupling between the Weyl and the Maxwell fields in cosmology has been analysed in~\cite{TDM,T1}, with the latter article arguing for a potentially very strong (resonant-type) amplification of large-scale magnetic fields. It should be noted, however that~\cite{T1} did not explicitly show the amplitude of the $B$-field to diverge, when the latter oscillated in tune with the gravitational waves. Here, we take a step further in that direction. We consider two scenarios and provide fully analytical (and gauge-invariant) treatments in both cases. The first operates in highly conductive environments and the second assumes poor electrical conductivity. In the former case, the gravito-magnetic interaction causes the superadiabatic amplification of the $B$-field on super-Hubble lengths. The standard adiabatic magnetic decay-rate ($B\propto a^{-2}$, where $a$ is the scale factor) slows down. In the second scenario, where the conductivity is very low, the Weyl-Maxwell coupling proceeds differently. There, the solutions show resonant magnetic amplification on all scales of interest. Driven by the gravitational waves, the $B$-field oscillates with an amplitude that is found to diverge when the wavelengths of the two sources coincide.

The objectives of the present article are (i) to demonstrate that the nature and the outcome of the gravito-magnetic interaction depends on the electric properties of the environment, since there is considerable confusion on this issue in (some of) the literature, and (ii) to show that Weyl curvature distortions can resonantly amplify the $B$-field. Mathematically speaking the latter means that the equations contain resonant solutions. The basic physics behind these gravito-magnetic resonances is relatively simple. Gravitational waves are known to trigger shear inhomogeneities that affect the magnetic evolution. In poorly conductive environments, where the $B$-field naturally oscillates, the aforementioned shear perturbations act as an external driving force. Forced vibrations, however, are well known to provide the physical stage for resonances to occur. It is not surprising then that there exist analytical solutions showing the magnetic amplitude to diverge when the Weyl and the Maxwell fields oscillate in tune. Realistically speaking, of course, what such diverging solutions indicate is an increased chance of a disproportionately strong amplification. This is also what our analysis suggests: that it is theoretically possible to substantially amplify cosmological $B$-fields by means of very weak Weyl-curvature distortions alone.

\section{The Weyl-Maxwell coupling in cosmology}\label{sW-MCC}
In a general spacetime, with metric $g_{ab}$ of signature ($-,+,+,+$), we define a timelike vector field $u^a={\rm d}x^a/{\rm d}\tau$ that is tangent to the worldlines of the fundamental observers (with $\tau$ representing their associated proper time). Together with the projector, $h_{ab}=g_{ab}+u_au_b$, the $u_a$-field introduces a unique 1+3 threading of the spacetime into time and space, while it decomposes every variable, every differential operator and every vector/tensor equation into their timelike and spacelike components (see~\cite{EvE} for details).

\subsection{Background dynamics}\label{ssBDs}
Consider a spatially flat FRW spacetime filled with a single barotropic fluid and permeated by a weak magnetic field ($\tilde{B}_a$). The latter has energy density well below that of the matter (i.e.~$\tilde{B}^2= \tilde{B}_a\tilde{B}^a\ll\rho$) to ensure that the symmetries and the evolution of the of the host universe remain unaffected. We may therefore ignore terms of order $\tilde{B}^2$ and the magnetic contribution to the background dynamics. The zero-order model is then determined by the standard set
\begin{equation}
H^2={1\over3}\,\rho\,, \hspace{10mm} \dot{H}= -H^2- {1\over6}\,\rho(1+3w) \hspace{10mm} {\rm and} \hspace{10mm} \dot{\rho}= -3(1+w)H\rho\,,  \label{bFEqs}
\end{equation}
where $w=p/\rho$ is the barotropic index of the fluid ($p$ is its isotropic pressure), $H=\dot{a}/a$ is the Hubble parameter and overdots indicate (proper) time derivatives.\footnote{Throughout this paper we use geometrised units with $8\pi G=1=c$.} At the same time,
\begin{equation}
\dot{\tilde{B}}_a= -2H\tilde{B}_a\,,  \label{bEvEqs}
\end{equation}
monitors the zero-order evolution of the (test) $\tilde{B}$-field~\cite{TB}.

Next, we will perturb the above background by allowing, among others, for gravitational-wave distortions. To first order, these are monitored by shear perturbations (see \S~\ref{ssG-MI} below), which couple to the magnetic field through the induction equation~\cite{TB}. Our aim is to study the effects of gravitational waves on the evolution of large-scale magnetic fields during the early (pre-recombination) stages of the expansion. The covariant expressions governing the gravito-magnetic interaction have been derived in~\cite{T1}, where the reader is referred to for the details. Here, we will simply present the relations and use them to study the Weyl-Maxwell coupling in cosmology. Further discussion on the formalism and its applications can be found in~\cite{TB}.

Before closing this section, it is worth noting that we follow the approach of~\cite{TDM,T1}, which also allows for a weak background (test) magnetic field. Alternatively, one could treat both the $\tilde{B}$-field and the gravitational waves as first-order perturbations and study their interaction at second order~\cite{FD}. The latter scheme maintains the purity of the FRW background, but only allows shear-magnetic terms into the equations. All other second-order couplings are excluded. As a result, when it comes to the evolution of the $B$-field, the two approaches are equivalent. The second-order magnetic wave equation of~\cite{FD} is the same with that of~\cite{T1} (compare Eq. (45) in~\cite{FD} -- see also comment immediately after expression (47) there -- with Eq.~(5) in~\cite{T1}, or with relations (\ref{ddotBa1}), (\ref{dBddot}) here).

\subsection{The gravito-magnetic interaction}\label{ssG-MI}
Information on propagating gravitational waves is encoded in the non-local component of the Riemann curvature tensor, which is commonly referred to as the Weyl field. The latter splits into an electric and a magnetic part, represented by the symmetric and trace-free tensors $E_{ab}$ and $H_{ab}$ respectively. The transverse components of these quantities describe gravitational waves, with their transversality guaranteeing that only the pure-tensor modes are accounted for and all scalar and vector perturbations have been switched off. In that case the magnetic Weyl component is directly related to the shear by means of the linear constraint~\cite{EvE}
\begin{equation}
H_{ab}= {\rm curl}\sigma_{ab}\,,  \label{Hcon}
\end{equation}
with ${\rm curl}S_{ab}=\varepsilon_{cd\langle a}{\rm D}^cS_{b\rangle}{}^d$ for any orthogonally projected, symmetric and trace-free tensor $S_{ab}$. All these mean that, to first order, we can monitor gravitational waves using the wave equation of the shear. On a spatially flat FRW background, the latter linearises to~\cite{T1,EvE}
\begin{equation}
\ddot{\sigma}_{ab}- {\rm D}^2\sigma_{ab}+ 5H\dot{\sigma}_{ab}+ {3\over2}\,(1-3w)H^2\sigma_{ab}=0\,,  \label{ddotsigma}
\end{equation}
where ${\rm D}^2={\rm D}_a{\rm D}^a$ is the orthogonally projected Laplacian operator.\footnote{In~\cite{FD} the shear was also used to describe gravitational waves at second order. Beyond the linear level, however, expression (\ref{Hcon}) no longer holds and the shear alone cannot (generally) monitor gravitational-wave distortions.}

The electromagnetic field obeys Maxwell's equations which combine to provide a set of wave equations, one for each component (see expressions (39) and (40) in~\cite{T2}). After all scalar and vector perturbations have been switched off, and given our weakly magnetised background, the wave formula of the $B$-field reduces to
\begin{equation}
\ddot{B}_a- {\rm D}^2B_a+ 5H\dot{B}_a+ 3(1-w)H^2B_a= 2\left(\dot{\sigma}_{ab}+2H\sigma_{ab}\right)\tilde{B}^b+ {\rm curl}\mathcal{J}_a\,.  \label{ddotBa1}
\end{equation}
The above monitors the gravito-magnetic interaction on a spatially flat FRW background at the lowest perturbative order~\cite{T1}. Note that $\tilde{B}_a$ is the background $B$-field (see \S~\ref{ssBDs}) and $\mathcal{J}_a$ the electric 3-current. This depends on the electric properties of the medium, which are encoded in Ohm's law. In the frame of the fluid (i.e.~that of the fundamental observers), the latter takes the covariant form
\begin{equation}
\mathcal{J}_a= \varsigma E_a\,,  \label{Ohm}
\end{equation}
where $\varsigma$ is the electric conductivity of the matter~\cite{G}. In poorly conductive environments, $\varsigma\rightarrow0$ and there are no currents. Then, we may ignore the last term in the right-hand side of (\ref{ddotBa1}). At the opposite end, where the conductivity is very high and $\varsigma\rightarrow \infty$, the electric field vanishes. In that case, $\mathcal{J}_a={\rm curl}B_a$ (e.g.~see~\cite{TB}) and expression (\ref{ddotBa1}) reduces to
\begin{equation}
\dot{B}_a+ 2HB_a= \sigma_{ab}\tilde{B}^b\,.  \label{dotBa}
\end{equation}
Not surprisingly, in the absence of its electric counterpart, the $B$-field no longer obeys a wave-like formula. Instead, the magnetic evolution is determined by the familiar induction equation~\cite{TB}.

\section{Gravitationally driven magnetic field}\label{sGDMF}
The significant differences between expressions (\ref{ddotBa1}) and (\ref{dotBa}) suggest that the electric properties of the cosmic medium, at the time and in the region the gravito-magnetic interaction occurs, are crucial. Here, we will mainly focus on the two limiting cases of very low and very high electrical conductivity. The intermediate case of finite resistivity will also be discussed in \S~\ref{ssFC}.

\subsection{Harmonic splitting}\label{ssHS}
Gravitational waves introduce shear inhomogeneities, which couple to the background magnetic field through Maxwell's formulae. The resulting gravito-magnetic stresses drive the evolution of the perturbed (the induced) $B$-field according to Eqs.~(\ref{ddotBa1}) or (\ref{dotBa}), depending on the conductivity of the environment. Our next step is to harmonically decompose these expressions. This will also provide a relation between the wavelengths of the agents involved. Splitting the zero-order field as $\tilde{B}_a= \tilde{B}_{(n)} \tilde{\mathcal{Q}}_a^{(n)}$, helps to assign a scale and a finite wavenumber ($n$) to it. Similarly, we may set $\sigma_{ab}= \Sigma_{k}\sigma_{(k)}\mathcal{Q}_{ab}^{(k)}$ and $B_a=\Sigma_{\ell}B_{(\ell)}\mathcal{Q}_a^{(\ell)}$, where $k$, $\ell$ are the associated wavenumbers and $\mathcal{Q}_a^{(\ell)}= \mathcal{Q}_{ab}^{(k)}\tilde{\mathcal{Q}}_{(n)}^b$ by construction~\cite{TDM,T1}. The three corresponding wavevectors satisfy the relation $\ell^2=\ell_a\ell^a=(n_a+k_a)(n^a+k^a)$. This immediately translates into
\begin{equation}
\ell= k\left[1+\left({n\over k}\right)^2 +2\left({n\over k}\right)\cos\varphi\right]^{1/2}\,,  \label{ell}
\end{equation}
with $0\leq\varphi\leq\pi$ representing the angle between the wavevector of the background magnetic field and that of the oncoming gravitational wave.

\subsection{Evolution equations}\label{ssEEs}
The harmonic decomposition of the previous section reduces the wave formulae (\ref{ddotsigma}) and (\ref{ddotBa1}) from partial to ordinary differential equations. Together with the decomposed counterpart of expression (\ref{dotBa}), the new set consists of
\begin{equation}
\ddot{\sigma}_{(k)}+ 5H\dot{\sigma}_{(k)}+ \left[{3\over2}\,(1-3w)H^2+\left({k\over a}\right)^2\right] \sigma_{(k)}= 0\,,  \label{dsigmaddot}
\end{equation}
\begin{equation}
\ddot{B}_{(\ell)}+ 5H\dot{B}_{(\ell)}+ \left[3(1-w)H^2+ \left({\ell\over a}\right)^2\right]B_{(\ell)}= 2\left(\dot{\sigma}_{(k)}+2H\sigma_{(k)}\right)\tilde{B}_{(n)}  \label{dBddot}
\end{equation}
and
\begin{equation}
\dot{B}_{(\ell)}+ 2HB_{(\ell)}= \tilde{B}_{(n)}\sigma_{(k)}\,.  \label{dBdot}
\end{equation}
The first two monitor the gravito-magnetic interaction in poorly conductive cosmological environments, while the set (\ref{dsigmaddot}), (\ref{dBdot}) does the same when the electric conductivity is high and the ideal magnetohydrodynamic (MHD) limit applies.

We proceed further by introducing the dimensionless, expansion normalised variables $\Sigma_{(k)}=\sigma_{(k)}/3H$ and $\mathcal{B}_{(\ell)}= B_{(\ell)}/3H$. Then, using conformal ($\eta$, with $\dot{\eta}=1/a$) instead of proper time, the system (\ref{dsigmaddot})-(\ref{dBdot}) recasts into
\begin{equation}
\Sigma^{\prime\prime}_{(k)}+ (1-3w)\left({a^{\prime}\over a}\right)\Sigma^{\prime}_{(k)}- \left\{{3[1+(2-3w)w]\over2}\left({a^{\prime}\over a}\right)^2 -k^2\right\}\Sigma_{(k)}= 0\,,  \label{eq:Sigma''}
\end{equation}
\begin{eqnarray}
\mathcal{B}_{(\ell)}^{\prime\prime}+ (1-3w)\left({a^{\prime}\over a}\right)\mathcal{B}_{(\ell)}^{\prime}- \left[{3(1-3w)w\over2}\left({a^{\prime}\over a}\right)^2 -\ell^2\right]\mathcal{B}_{(\ell)}= \nonumber\\ ={2a_0^2\tilde{B}_0^{(n)}\over a} \left[\Sigma_{(k)}^{\prime}+{1-3w\over2}\left({a^{\prime}\over a}\right)\Sigma_{(k)}\right]  \label{eq:cB''}
\end{eqnarray}
and
\begin{equation}
\mathcal{B}_{(\ell)}^{\prime}+ {1\over2}\,(1-3w)\left({a^{\prime}\over a}\right)\mathcal{B}_{(\ell)}= {a_0^2\tilde{B}_0^{(n)}\over a}\, \Sigma_{(k)}\,,  \label{eq:cB'}
\end{equation}
where primes indicate conformal-time derivatives and the zero suffix marks the beginning of the gravito-magnetic interaction. Also, in deriving Eqs.~(\ref{eq:cB''}) and (\ref{eq:cB'}) we have used the evolution law, $\tilde{B}_{(n)}\propto a^{-2}$, of the background magnetic field (see Eq.~(\ref{bEvEqs}) in \S~\ref{ssBDs}). As before, the set (\ref{eq:Sigma''}), (\ref{eq:cB''}) applies to poorly conductive environments, while Eqs.~(\ref{eq:Sigma''}) and (\ref{eq:cB'}) operate at the ideal-MHD limit. The advantage of the new notation is that, when evaluated in the radiation era, expressions (\ref{eq:Sigma''})-(\ref{eq:cB'}) take a very simple and compact form that is also straightforward to solve analytically. In addition, during the radiation era, the expansion normalised $\mathcal{B}$-field satisfies the criteria for gauge invariance (see \S~\ref{sGAMF} below).

Mathematically speaking, expression (\ref{eq:cB''}) is a wave formula and (\ref{eq:cB'}) is a first-order differential equation. Physically, this difference reflects the absence of electric fields in highly conductive environments. What is most important, however, is that in (\ref{eq:cB''}) -- as well as in (\ref{dBddot}) -- the gravitational waves drive the magnetic oscillation. Forced oscillations, however, are known to provide the natural physical stage for resonances to occur. In other words, expressions (\ref{dBddot}) and (\ref{eq:cB''}) open the theoretical possibility of gravito-magnetic resonances on cosmological scales.

\section{Gravitationally amplified magnetic field}\label{sGAMF}
The galactic dynamo requires seed fields coherent on a minimum comoving scale of roughly 10~Kpc. In typical cosmological models such scales have been outside the Hubble horizon for a long time. Therefore, in order to achieve astrophysically interesting results, we should consider lengths close and beyond the Hubble radius.

\subsection{Resonant amplification}\label{ssRA}
Inflation has long been known to provide a mechanism that produces electromagnetic fields and gravitational waves with super-horizon correlations. Also, throughout the inflationary phase, the universe is a very poor electrical conductor. Although the conductivity is expected to grow quickly with the onset of the radiation era, causality must confine its effects (i.e.~the currents that will eliminate the electric fields) well inside the horizon. Near and beyond the Hubble radius, the gravito-magnetic interaction will still proceed in line with Eqs.~(\ref{eq:Sigma''}) and (\ref{eq:cB''}).

In the radiation era the expansion normalised variables introduced in \S~\ref{ssEEs} have the double advantage of simplifying the equations and also of freeing the results from gauge-related ambiguities. The gauge issue concerns the magnetic field, since the shear vanishes at the zero-order level and all the shear-related variables are therefore gauge independent. The presence of a nonzero background magnetic field, on the other hand, means that the magnetic-related variables are generally gauge dependent. During the radiation era, however, the scalar $B_{(\ell)}/3H$ remains constant to zero order and therefore satisfies the Stewart \& Walker criteria for gauge invariance~\cite{SW}. As a result, when radiation dominates the energy density of the universe, the expansion normalised variable $\mathcal{B}_{(\ell)}$ is a gauge-invariant linear perturbation.

With these in mind, we apply the system (\ref{eq:Sigma''}), (\ref{eq:cB''}) to a radiation dominated almost-FRW universe. Setting $w=1/3$, $a=a_0^2H_0\eta$ and $a^{\prime}/a=1/\eta$, Eq.~(\ref{eq:Sigma''}) reduces to
\begin{equation}
\Sigma^{\prime\prime}_{(k)}- \left({2\over\eta^2}-k^2\right)\Sigma_{(k)}= 0\,,  \label{eq:rSigma''}
\end{equation}
and accepts the oscillatory solution
\begin{equation}
\Sigma_{(k)}= \mathrm{C}_1\left[\sin(k\eta)+{\cos(k\eta)\over k\eta}\right]+ \mathrm{C}_2\left[\cos(k\eta)- {\sin(k\eta)\over k\eta}\right]\,.  \label{eq:rSigma1}
\end{equation}
Keeping only the dominant $\Sigma$-mode, we may rewrite the above as\footnote{Including the subdominant $\Sigma$-mode in (\ref{eq:rSigma2}) simply adds decaying oscillatory terms to the right-hand side of solutions (\ref{eq:rcB''2}) and (\ref{eq:rcB}), without changing neither the nature of these expressions nor the overall results.}
\begin{eqnarray}
\Sigma_{(k)}&=& \left[\Sigma_0^{(k)}\sin(k\eta_0)+{1\over k}\, \Sigma_0^{\prime\,(k)}\cos(k\eta_0)\right]\sin(k\eta) \nonumber\\ &&+\left[\Sigma_0^{(k)}\cos(k\eta_0)-{1\over k}\, \Sigma_0^{\prime\,(k)}\sin(k\eta_0)\right]\cos(k\eta)\,,  \label{eq:rSigma2}
\end{eqnarray}
with the zero suffix indicating the onset of the gravito-magnetic interaction -- which for our purposes coincides with the beginning of the radiation era.

At the same time and on the same scales, the gravitationally driven magnetic field obeys the wave formula
\begin{equation}
\mathcal{B}_{(\ell)}^{\prime\prime}+ \ell^2\mathcal{B}_{(\ell)}= {6\over\eta}\,\tilde{\mathcal{B}}_0^{(n)} \Sigma_{(k)}^{\prime}\,,  \label{eq:rcB''1}
\end{equation}
with $\tilde{\mathcal{B}}_{(n)}=\tilde{B}_{(n)}/3H$ and $\Sigma_{(k)}$ given by solution (\ref{eq:rSigma2}). Differentiating the latter with respect to conformal time and substituting the result into the right-hand side of (\ref{eq:rcB''1}), we arrive at
\begin{eqnarray}
\mathcal{B}_{(\ell)}^{\prime\prime}+ \ell^2\mathcal{B}_{(\ell)}&=&
-{6k\over\eta}\,\tilde{\mathcal{B}}_0^{(n)} \left[\Sigma_0^{(k)}\cos(k\eta_0)-{1\over k}\, \Sigma_0^{\prime\,{(k)}}\sin(k\eta_0)\right]\sin(k\eta) \nonumber\\ &&+{6k\over\eta}\,\tilde{\mathcal{B}}_0^{(n)} \left[\Sigma_0^{(k)}\sin(k\eta_0)+{1\over k}\, \Sigma_0^{\prime\,{(k)}}\cos(k\eta_0)\right]\cos(k\eta)\,.  \label{eq:rcB''2}
\end{eqnarray}
The above traces the evolution of the expansion normalised, gravitationally driven magnetic field during the radiation epoch of an almost-FRW universe near and outside the Hubble horizon. The solution of Eq.~(\ref{eq:rcB''2}) reads\footnote{Mathematically speaking, Eqs.~(\ref{eq:rcB''2}) and (\ref{eq:rcB}) hold on all scales, although in this section the `operational' domain of these expressions lies near and outside the Hubble radius. This does not always need to be the case however. As we will see in \S~\ref{ssFC}, Eq.~(\ref{eq:rcB''2}) and solution (\ref{eq:rcB}) can also extend to subhorizon lengths. }
\begin{eqnarray}
\mathcal{B}_{(\ell)}&=& \pm{3\tilde{\mathcal{B}}_0^{(n)}\sqrt{k^2\Sigma_0^2 +\Sigma_0^{\prime\,2}}\over\ell}\,{\rm Si}[(\ell\mp k)\eta] \sin\left[(\ell\eta\mp k\eta_0)\mp\phi\right] \nonumber\\ &&\pm{3\tilde{\mathcal{B}}_0^{(n)}\sqrt{k^2\Sigma_0^2 +\Sigma_0^{\prime\,2}}\over\ell}\,{\rm Ci}[(\ell\mp k)\eta] \cos\left[(\ell\eta\mp k\eta_0)\mp\phi\right]\,,  \label{eq:rcB}
\end{eqnarray}
where $\tan\phi=\Sigma_0^{\prime\,(k)}/k\Sigma_0^{(k)}$ (see Appendix for the details). Also, ${\rm Si}(x)$, ${\rm Ci}(x)$ represent the sine and the cosine integral functions respectively~\cite{AS}. Therefore, driven by the gravitational waves, the expansion normalised, gravitationally driven magnetic field oscillates with an amplitude that depends on the initial conditions. As typical for forced oscillations, the amplitude also depends on the wavenumbers of the agents involved. Given that
\begin{equation}
\lim_{x\rightarrow0}{\rm Si}(x)=0 \hspace{15mm} {\rm and} \hspace{15mm} \lim_{x\rightarrow0}{\rm Ci}(x)=-\infty\,,  \label{SiCi}
\end{equation}
we deduce that when the two waves are in resonance (namely as $k\rightarrow\ell$) the amplitude of the magnetic field increases arbitrarily. Note that, at least theoretically speaking, resonant effects do not depend on the strength of the driving source. In this respect, the magnetic amplification seen in solution (\ref{eq:rcB}) requires only the mere presence of gravitational waves and is rather insensitive to the amount of energy stored in them.

We may also express solution (\ref{eq:rcB}) in terms of the actual (instead of the expansion normalised) variables and with respect to proper (instead of conformal) time. To do that recall that $\mathcal{B}_{(\ell)}=B_{(\ell)}/3H$, $\Sigma_{(k)}=\sigma_{(k)}/3H$ and $\tilde{\mathcal{B}}_{(n)}=\tilde{B}_{(n)}/3H$ by definition, while $\Sigma^{\prime}_{(k)}=a\dot{\sigma}_{(k)}/3H$ after keeping only the dominant shear mode. Also, $H=1/2t$ and $\eta=\sqrt{t}/a_0H_0\sqrt{t_0}$ during the radiation era. Then, after some lengthy but straightforward algebra, we arrive at
\begin{eqnarray}
B_{(\ell)}&=& \pm{\tilde{B}_0^{(n)}\sqrt{k^2\sigma_0^2 +a_0^2\dot{\sigma}_0^2}\over\ell H_0}\left({t_0\over t}\right) {\rm Si}\left({\ell\mp k\over a_0H_0}\sqrt{t\over t_0}\right) \sin\left[{\ell\over a_0H_0}\left(\sqrt{t\over t_0} \mp{k\over\ell}\right)\mp\phi\right] \nonumber\\ &&\pm{\tilde{B}_0^{(n)}\sqrt{k^2\sigma_0^2 +a_0^2\dot{\sigma}_0^2}\over\ell H_0}\left({t_0\over t}\right) {\rm Ci}\left({\ell\mp k\over a_0H_0}\sqrt{t\over t_0}\right) \cos\left[{\ell\over a_0H_0}\left(\sqrt{t\over t_0} \mp{k\over\ell}\right)\mp\phi\right]\,, \label{eq:rB}
\end{eqnarray}
where now $\tan\phi=a\dot{\sigma}^{(k)}_0/k\sigma^{(k)}_0$. The above shows that the gravitationally driven (actual) magnetic field decays adiabatically (i.e.~$B_{(\ell)}\propto t^{-1}\propto a^{-2}$) like its background counterpart. However, as in solution (\ref{eq:rcB}), the presence of the cosine integral function ${\rm Ci} [(\ell-k)\sqrt{t}/a_0H_0\sqrt{t_0}]$ in the right-hand side of (\ref{eq:rB}) guarantees that the amplitude of the $B$-field diverges when the two sources are in resonance (i.e.~for $k\rightarrow\ell$). It goes without saying that solution (\ref{eq:rB}) can be obtained directly from the system (\ref{dsigmaddot}), (\ref{dBddot}). In particular, solving Eq.~(\ref{dsigmaddot}), keeping the dominant shear mode and then substituting into (\ref{dBddot}) leads to expression (\ref{eq:rB}). To the best of our knowledge, this is the first time the amplitude of cosmological (or astrophysical) magnetic fields has been found to diverge, as a result of their interaction with gravitational-wave distortions.

Solutions (\ref{eq:rcB}) and  (\ref{eq:rB}) describe one type of resonance between the Weyl and the Maxwell fields, but there are other possibilities as well. In each individual case the characteristics of the resonance are decided by the shape of Eq.~(\ref{eq:rcB''1}) and by the form of the driving term in the right-hand side of that expression. Suppose that the source term has time-independent amplitude. This can happen, for example, when the background magnetic field decays as $\eta^{-1}$ (like many of the $B$-fields proposed in~\cite{TW}-\cite{TK}), while the expansion normalised shear is still given by (\ref{eq:rSigma2}). In that case Eq.~(\ref{eq:rcB''1}) accepts the solution
\begin{equation}
\mathcal{B}_{(\ell)}= {2a_0\tilde{B}_0^{(n)} \sqrt{k^2\Sigma_0^2+\Sigma_0^{\prime\,2}}\over k^2-\ell^2} \,\sin[k(\eta-\eta_0)-\phi]\,,  \label{eq:alrcB}
\end{equation}
which also ensures arbitrary growth for the amplitude of the gravitationally driven $\mathcal{B}$-field as $k\rightarrow\ell$. It is therefore plausible to argue that a greater variety of physical environments could support analogous effects. It is also very likely, however, that only a small fraction of these cases will be analytically tractable. Overall, our analysis opens the theoretical possibility for gravity to act, through its long-range component, as a very efficient magnetic dynamo on cosmological scales and long before the onset of structure formation.

In retrospect, the results reported in this section should not come as a surprise. After all, we were dealing with forced oscillations, where the Weyl part of the gravitational field was the driving agent. Forced oscillations have long been known to provide the physical stage where resonances naturally occur. In this sense, our study simply shows that resonances can also take place between cosmological gravitational waves and large-scale magnetic fields.

Finally, before closing this section, we should underline that solution (\ref{eq:rB}) and the resulting gravito-magnetic resonance could have been obtained in~\cite{FD} as well. For example, solve the linearised version of Eq.~(46) -- the numbering is that of~\cite{FD} -- during the radiation era and on all scales. Then, substitute the wave solution of (46) into Eq.~(45) -- the numbering is still that of~\cite{FD} -- and solve again. The result would have led to expression (\ref{eq:rB}) above.

\subsection{Superadiabatic amplification}\label{ssSA}
Let us now turn to the ideal-MHD limit. The main objective is to illustrate the qualitative change in the nature of the gravito-magnetic interaction caused by the different electric properties of the cosmic medium. To some extent this is to be expected, given the significant differences between Eqs.~(\ref{ddotBa1}) and (\ref{dotBa}) -- or between expressions (\ref{eq:cB''}) and (\ref{eq:cB'}). To keep things simple, we assume that the magnetic field is frozen-in with the matter on all scales. This is possible within, say, the Eddington-Lemaitre or the `emergent-universe' scenarios~\cite{EM}, where the cosmos evolves from a past-eternal static state. Then, in the frame of the fluid, the gravito-magnetic interaction is monitored by Eqs.~(\ref{eq:Sigma''}) and (\ref{eq:cB'}). This means that during the radiation era the gravitationally driven magnetic field evolves according to
\begin{equation}
\mathcal{B}_{(\ell)}^{\prime}= {3\over\eta}\,\tilde{\mathcal{B}}_0^{(n)}\Sigma_{(k)}\,.  \label{eq:MHDrcB'}
\end{equation}
with $\tilde{\mathcal{B}}_{(n)}=\tilde{B}_{(n)}/3H$ and $\Sigma_{(k)}$ given by (\ref{eq:rSigma1}). As before, the zero suffix indicates the beginning of the interaction. The form of expression (\ref{eq:MHDrcB'}) leads to the first qualitative conclusion. In particular, the fact that the $B$-field no longer obeys a wave equation means that gravito-magnetic resonances are not possible in highly conductive environments (see also \S~VIB in~\cite{T1}).

To demonstrate the effect of the Weyl field at the ideal MHD limit, it suffices to consider super-Hubble lengths and replace solution (\ref{eq:rSigma1}) with its long-wavelength approximation, namely
\begin{equation}
\Sigma_{(k)}= {1\over3}\left(\Sigma_0^{(k)}+\eta_0\Sigma_0^{\prime\,(k)}\right) \left({\eta\over\eta_0}\right)^2+ {1\over3}\left(2\Sigma_0^{(k)}-\eta_0\Sigma_0^{\prime\,(k)}\right) \left({\eta_0\over\eta}\right)\,.  \label{eq:rlSigma}
\end{equation}
Substituting this into Eq.~(\ref{eq:MHDrcB'}), integrating and setting the integration constant to zero for simplicity, leads to
\begin{equation}
\mathcal{B}_{(\ell)}= {1\over2}\,\tilde{\mathcal{B}}_0^{(n)} \left(\Sigma_0^{(k)}+\eta_0\Sigma_0^{\prime\,(k)}\right) \left(\eta\over\eta_0\right)^2- \tilde{\mathcal{B}}_0^{(n)} \left(2\Sigma_0^{(k)}-\eta_0\Sigma_0^{\prime\,(k)}\right) \left({\eta_0\over\eta}\right)\,,  \label{eq:MHDrcB}
\end{equation}
on all super-Hubble lengths. Finally, recalling that $\mathcal{B}_{(\ell)}=B_{(\ell)}/3H$, that $\tilde{\mathcal{B}}_{(n)}=\tilde{B}_{(n)}/3H$ and that $H=1/a_0^2H_0\eta^2$ throughout the radiation era -- which also implies that $a_0H_0\eta_0=1$ at the time, we arrive at
\begin{equation}
B_{(\ell)}= {1\over2}\,\tilde{B}_0^{(n)} \left(\Sigma_0^{(k)}+\eta_0\Sigma_0^{\prime\,(k)}\right)- \tilde{B}_0^{(n)} \left(2\Sigma_0^{(k)}-\eta_0\Sigma_0^{\prime\,(k)}\right) \left({\eta_0\over\eta}\right)^3\,.  \label{eq:MHDrB}
\end{equation}

Following (\ref{eq:MHDrB}), prior to equipartition, the gravitationally driven magnetic field no longer decays adiabatically. Instead, the dominant magnetic mode remains constant, which impies a superadiabatic type of amplification for the $B$-field on all super-Hubble scales due to Weyl curvature effects. This means that magnetic fields that cross inside the Hubble radius around the time of matter-radiation equilibrium have magnitudes equal to
\begin{equation}
B_{H}^{(\ell)}= {1\over2} \left(\Sigma_0^{(k)}+\eta_0\Sigma_0^{\prime\,(k)}\right) \tilde{B}_0^{(n)}\,.  \label{eq:MHDrBf}
\end{equation}
Thus, the strength of the residual $B$-field depends on that of its background counterpart at the onset of the interaction and on the relative shear anisotropy at the same time. In particular, the final outcome is decided by the values of $\Sigma_0^{(k)}$ and $\Sigma_0^{\prime\,(k)}$, of which the first (at least) is expected to be very small. Therefore, unless $\eta_0\Sigma_0^{\prime}\gg\Sigma_0$ -- or the shear anisotropy is larger than anticipated -- this effect is unlikely to produce magnetic fields of astrophysical importance.

\subsection{Finite conductivity}\label{ssFC}
So far, we have focussed on the two limiting cases of very low and very high electrical conductivity. For completeness, let us take a brief qualitatively look at the intermediate scenario of finite resistivity. Assuming that the conductivity coefficient has a nearly homogeneous spatial distribution (i.e.~setting ${\rm D}_a \varsigma\simeq0$), Eqs.~(\ref{ddotBa1}) and (\ref{Ohm}) combine to
\begin{equation}
\ddot{B}_a- {\rm D}^2B_a+ 5\left(1+{1\over5}\,\tilde{\varsigma}\right)H\dot{B}_a+ 3\left(1-w+{2\over3}\,\tilde{\varsigma}\right)H^2B_a= 2\left[\dot{\sigma}_{ab}+2\left(1+{1\over4}\, \tilde{\varsigma}\right)H\sigma_{ab}\right]\tilde{B}^b\,,  \label{eq:ddotBa2}
\end{equation}
where $\tilde{\varsigma}=\varsigma/H$ is the expansion normalised conductivity scalar (see \S~VIA in~\cite{T1} for more details on the derivation). As long as $\tilde{\varsigma}\ll1$, the above reduces to Eq.~(\ref{dBddot}) and the Weyl-Maxwell coupling proceeds as in poorly conductive mediums irrespective of the scale.  Therefore, gravito-magnetic resonances, like those discussed in \S~\ref{ssRA} before, can also occur on subhorizon lengths, as long as $\tilde{\varsigma}\ll1$ there

When the ratio $\varsigma/H$ is large, we can no longer ignore the role and the effects of the electric conductivity (on sub-Hubble scales). Then, expression (\ref{eq:ddotBa2}) assumes the form
\begin{equation}
\ddot{B}_a- {\rm D}^2B_a+ \tilde{\varsigma}H\dot{B}_a+ 2\tilde{\varsigma}H^2B_a= 2\left(\dot{\sigma}_{ab}+{1\over2}\, \tilde{\varsigma}H\sigma_{ab}\right)\tilde{B}^b\,,  \label{ddotBa3}
\end{equation}
and the gravito-magnetic interaction depends largely on the electric properties of the medium. Situations like this require a model for the conductivity of the cosmic plasma. Note that the above relation applies to cosmological environments of significant but still finite conductivity and one should not simply extrapolate it to the ideal-MHD limit (where $\varsigma\rightarrow\infty$). There, the electric fields vanish (see Eq.~(\ref{Ohm})) and one should instead use Eqs.~(\ref{dBdot}) or (\ref{eq:cB'}).

\section{Discussion}
The widespread presence of magnetic fields in the universe is a hard observational fact. From the nearby planets and stars, all the way to distant galaxies, galaxy clusters and remote high-redshift proto-galactic clouds, the magnetic presence has been repeatedly verified. The origin of the large-scale $B$-fields, however, is still illusive. Typically, magnetic fields with coherence scales large enough to sustain the galactic dynamo are too weak, while those of adequate strength span very small lengths. Inflation can naturally produce large-scale correlations and it has long been seen as the strongest candidate for early-time magnetogenesis. Nevertheless, $B$-fields that survive the exponential phase are typically too weak to seed the dynamo, unless classical electromagnetism is modified or FRW hosts with non-Euclidean spatial geometry are involved.

Geometry could provide a solution to the magnetic-strength problem through Weyl-curvature distortions as well. Gravitational waves, which are readily produced during the inflationary era,  interact with magnetic fields via the shear anisotropy that they introduce. The outcome of the coupling between the Weyl and the Maxwell fields depends, among others, on the electric properties of the medium that fills the universe. In highly conductive environments, where the currents keep the $B$-field frozen-in with the fluid, the gravito-magnetic interaction slows down the standard adiabatic magnetic decay. Looking at the radiation era, we found a superadiabatic-type of amplification. The strength of the field remained constant, instead of dropping in accord with the usual $a^{-2}$-law. Nevertheless, the overall magnetic growth is likely to be small, unless the amount of the initial shear anisotropy is higher than expected.

The Weyl-Maxwell coupling looks much more promising in poorly conductive environments, where the gravitational waves drive the magnetic oscillations. Forced oscillations are well known to provide the natural environment for resonances to occur. It is not surprising then that our analysis has found resonances between gravitational waves and magnetic fields. During the radiation era, in particular, the amplitude of the gravitationally driven $B$-field was found to diverge when the wavenumbers of the two sources coincided. Physically speaking, what resonant solutions mean is that very weak vibrations can have disproportionately strong effects. This is also what our analysis indicates: that very weak gravitational waves can strongly amplify cosmological magnetic fields.

Typically, the mechanism operates near and beyond the Hubble horizon. Smaller scales can also provide a favourable environment, as long as $\tilde{\varsigma}= \varsigma/H\ll1$ there (see Eq.~(\ref{eq:ddotBa2}) in \S~\ref{ssFC}). Given that inflation produces gravitational waves and magnetic fields essentially on all lengths, the chances of resonances to occur (i.e.~the probability of their wavelengths coinciding) should be relatively high. On causality grounds, the most likely time for the amplification to occur is at horizon crossing. Well inside the Hubble radius, the currents should eliminate the electric fields and freeze the magnetic component in, thus changing the nature of the interaction. Naively, one should also expect a similar amount of strengthening for every magnetic mode passing through the horizon. This favours a scale-invariant spectrum for essentially all the large-scale $B$-fields in the universe. The latter appears to be in qualitative agreement with the observations, which show magnetic fields of similar ($\mu$G-order) strength in nearby galaxies as well as in remote proto-galactic structures.

Finally, we should point out that cosmology is probably not the only venue where gravito-magnetic resonances can occur. In principle, analogous effects could also take place in astrophysical environments. After all, the nature and the basic physics of the interaction remain the same. With this in mind, it might be worth investigating whether (and under what conditions) the coupling between astrophysical gravitational waves and magnetic/electromagnetic fields could lead to the resonant amplification of the latter.\\

The author thanks Axel Brandenburg, Ruth Durrer and Elisa Fenu for helpful discussions.

\section*{Appendix}
Given the key role of solution (\ref{eq:rcB}), we provide some additional technical information and intermediate steps for the interested reader who wishes to recover the above named expression. To begin with, it is relatively straightforward to verify that Eq.~(\ref{eq:rcB''2}) can be solved analytically on all scales. The solution describes forced oscillation and takes the form
\begin{eqnarray}
\mathcal{B}_{(\ell)}&=& \mathcal{C}_1\sin(\ell\eta)+ \mathcal{C}_2\cos(\ell\eta) \nonumber\\ &&+{3k\tilde{\mathcal{B}}_0^{(n)}\Sigma_0^{(k)}\over\ell} \left[\cos(k\eta_0)\sin(\ell\eta)-\sin(k\eta_0)\cos(\ell\eta)\right] {\rm Si}\left[(\ell-k)\eta\right] \nonumber\\ &&-{3\tilde{\mathcal{B}}_0^{(n)}\Sigma_0^{\prime\,(k)}\over\ell} \left[\sin(k\eta_0)\sin(\ell\eta)+\cos(k\eta_0)\cos(\ell\eta)\right] {\rm Si}\left[(\ell-k)\eta\right]  \nonumber\\ &&+{3k\tilde{\mathcal{B}}_0^{(n)}\Sigma_0^{(k)}\over\ell} \left[\sin(k\eta_0)\sin(\ell\eta)+\cos(k\eta_0)\cos(\ell\eta)\right] {\rm Ci}\left[(\ell-k)\eta\right]  \nonumber\\ &&+{3\tilde{\mathcal{B}}_0^{(n)}\Sigma_0^{\prime\,(k)}\over\ell} \left[\cos(k\eta_0)\sin(\ell\eta)-\sin(k\eta_0)\cos(\ell\eta)\right] {\rm Ci}\left[(\ell-k)\eta\right]  \nonumber\\ &&-{3k\tilde{\mathcal{B}}_0^{(n)}\Sigma_0^{(k)}\over\ell} \left[\cos(k\eta_0)\sin(\ell\eta)+\sin(k\eta_0)\cos(\ell\eta)\right] {\rm Si}\left[(\ell+k)\eta\right]  \nonumber\\ &&+{3\tilde{\mathcal{B}}_0^{(n)}\Sigma_0^{\prime\,(k)}\over\ell} \left[\sin(k\eta_0)\sin(\ell\eta)-\cos(k\eta_0)\cos(\ell\eta)\right] {\rm Si}\left[(\ell+k)\eta\right]  \nonumber\\ &&+{3k\tilde{\mathcal{B}}_0^{(n)}\Sigma_0^{(k)}\over\ell} \left[\sin(k\eta_0)\sin(\ell\eta)-\cos(k\eta_0)\cos(\ell\eta)\right] {\rm Ci}\left[(\ell+k)\eta\right]  \nonumber\\ &&+{3\tilde{\mathcal{B}}_0^{(n)}\Sigma_0^{\prime\,(k)}\over\ell} \left[\cos(k\eta_0)\sin(\ell\eta)+\sin(k\eta_0)\cos(\ell\eta)\right] {\rm Ci}\left[(\ell+k)\eta\right]\,,  \label{eq:app1}
\end{eqnarray}
where $\mathcal{C}_{1,2}$ are the integration constants, while ${\rm Si}(x)$ and ${\rm Ci}(x)$ are the sine and the cosine integral functions respectively. Also, primes indicate conformal time derivatives and the zero suffix corresponds to the onset of the gravito-magnetic interaction, which for our purposes coincides with the beginning of the radiation era. Since the gravitationally induced magnetic field vanishes in the absence of shear perturbations, we may set the aforementioned integration constants to zero. Then, using standard trigonometry, solution (\ref{eq:app1}) reduces to the compact expression
\begin{eqnarray}
\mathcal{B}_{(\ell)}&=& \pm{3\tilde{\mathcal{B}}_0^{(n)}\over\ell} \left[k\Sigma_0^{(k)}\sin(\ell\eta\mp k\eta_0) \mp\Sigma_0^{\prime\,(k)}\cos(\ell\eta\mp k\eta_0)\right] {\rm Si}\left[(\ell\mp k)\eta\right] \nonumber\\ &&\pm{3\tilde{\mathcal{B}}_0^{(n)}\over\ell} \left[k\Sigma_0^{(k)}\cos(\ell\eta\mp k\eta_0) \pm\Sigma_0^{\prime\,(k)}\sin(\ell\eta\mp k\eta_0)\right] {\rm Ci}\left[(\ell\mp k)\eta\right] \,.  \label{eq:app2}
\end{eqnarray}
Finally, setting $\tan\phi=\Sigma_0^{\prime\,(k)}/k\Sigma_0^{(k)}$ and employing some relatively lengthy but straightforward algebra, we arrive at solution (\ref{eq:rcB})
\begin{eqnarray}
\mathcal{B}_{(\ell)}&=& \pm{3\tilde{\mathcal{B}}_0^{(n)}\sqrt{k^2\Sigma_0^2 +\Sigma_0^{\prime\,2}}\over\ell}\,{\rm Si}[(\ell\mp k)\eta] \sin\left[(\ell\eta\mp k\eta_0)\mp\phi\right] \nonumber\\ &&\pm{3\tilde{\mathcal{B}}_0^{(n)}\sqrt{k^2\Sigma_0^2 +\Sigma_0^{\prime\,2}}\over\ell}\,{\rm Ci}[(\ell\mp k)\eta] \cos\left[(\ell\eta\mp k\eta_0)\mp\phi\right]\,.  \label{eq:app3}
\end{eqnarray}
Note that, for simplicity, we have suppressed the wavenumbers ($k$) of the $\Sigma_0$ modes inside the square roots.

\end{document}